\documentclass[10pt, conference, a4paper]{IEEEtran}
\usepackage{graphicx}
\usepackage{amsmath}
\usepackage{amsfonts}
\usepackage{algpseudocode}
\usepackage{amsmath}
\usepackage{balance}
\usepackage[linesnumbered,algoruled,boxed,lined]{algorithm2e}
\makeatletter
\renewcommand\paragraph{\@startsection{paragraph}{4}{\z@}%
    {1.5ex plus .2ex minus .3ex}%
    {-0em}%
    {\normalsize\bf}}
\makeatother
\makeatletter
\newcommand{\removelatexerror}{\let\@latex@error\@gobble}
\makeatother
\IEEEoverridecommandlockouts

\title{TuRF: Fast Data Collection for Fingerprint-based Indoor Localization~\thanks{The first two authors made equal contributions to the work.}}
\author{\IEEEauthorblockN{Chenhe Li, Qiang Xu, Zhe Gong and  Rong Zheng}
\IEEEauthorblockA{Department of Computing and Software \\
McMaster University\\
Hamilton, ON, Canada\\
Email: \{{\it lic54, xuq22, gongz13, rzheng}\}@mcmaster.ca}
}

\begin{document}

\maketitle
\begin{abstract}
Many infrastructure-free indoor positioning systems rely on fine-grained location-dependent fingerprints to train models for localization. The site survey process to collect fingerprints is laborious and is considered one of the major obstacles to deploying such systems. In this paper, we propose TuRF, a fast path-based fingerprint collection mechanism for site survey. We demonstrate the feasibility to collect fingerprints for indoor localization during walking along predefined paths. A step counter is utilized to accommodate the variations in walking speed. Approximate location labels inferred from the steps are then used to train a Gaussian Process regression model. Extensive experiments show that TuRF can significantly reduce the required time for site survey, without compromising the localization performance. 
\end{abstract}
\section{Introduction}
Despite the fact that people spend majority of their time indoor, indoor positioning systems (IPS) only have limited success due to the lack of pervasive infrastructural support, and the desire to keep user devices as simple as possible. One major category of indoor localization solutions utilize location-dependent fingerprints (e.g. received signal strength (RSS) of WiFi, magnetic, luminous conditions.) to estimate indoor locations~\cite{Yang2012,prince2012two,xuzhengubicomp15}. Generally, these methods work in two stages: training and operational stages. In the training stage, comprehensive site survey is conducted to record the fingerprints at targeted locations. In the operational stage, when a user submits a location query with her current fingerprints, a localization server computes and returns the estimated location. 

Site survey for fingerprint-based localization is a laborious process and needs to be done repeatedly in case of changes in the environment and infrastructure. Take the authors' work space as an example. The area of the main corridor is about $500 m^2$. If the area is divided into $1.2m \times 1.2m$ grids, and 1 minute is spent per grid point -- a very conservative estimate for WiFi fingerprints, the whole process takes about 4.5 hours. This has not taken into account the amount of time to measure and mark the grid point locations. To expedite the site survey process, several researchers have proposed to leverage mobile crowdsensing to collect location-dependent fingerprints~\cite{Rai2012,Yang2012}. While this approach is attractive, it suffers from the problems of noisy data, poor coverage and possibly frauds~\cite{Xu:2016:MobiBee}. 

In this work, we develop TuRF, a fast fingerprint collection method, where users walk along predefined paths and record fingerprints using mobile phones. User locations along the paths are inferred through step counting to accommodate variations in walking speed. In contrast to traditional approaches that take multiple WiFi scans\footnote{One scan is defined as collecting the RSS data from all visible access points at one location.} at selected locations and use the average RSS values to train a regression model, we show that instantaneous RSS values collected at moderate walking speed can in fact be utilized to achieve comparable performance with significantly less time needed. We adopt Gaussian Process (GP) to train models for localization. Since the RSSes of different access points are observed at different locations during walking, the Gaussian process regression model for each access point is trained separately and merged afterwards for localization. Magnetic fingerprints are further incorporated to improve localization accuracy. Real world experiments show that the proposed path-based data collection method can be 9 times faster than traditional point-based method, without compromising the localization performance. Our main contributions are thus two-fold:
\begin{itemize}
\item {\it A fast fingerprint collection mechanism for indoor localization:} We propose a path-based mechanism to accelerate site survey. Step counting is utilized to accommodate the variations in walking speed. To make full use of fingerprints collected during walking, GP is employed to train a regression model for each access point separately. Maximum Likelihood Estimation method is then adopted for location estimation in the online phase.  

\item {\it A walking speed recommendation for path-based fingerprint collection:} The selection of walking speed is an important operation parameter. Walking too slowly lengthens the site survey time; walking too fast, on the other hand, results in insufficient training data. From extensive experiments, we come up with guidelines on selecting a proper  walking speed. 

\end{itemize}

The rest of this paper is organized as follows. A summary of the related work is given in Section~\ref{sect:relate}. In Section~\ref{sect:overview} we give a high level overview of the proposed solution approach.  Details are provided on the mechanism for fast data collection and regression model training in Section~\ref{sect:proposed}. Experimental results are presented in Section~\ref{sect:exp}. Finally, we conclude the paper and outline a list of future work in Section~\ref{sect:conclude}.

\section{Related Work}
\label{sect:relate}
Indoor positioning has received much attention in recent years. Existing solutions mainly fall into two categories: infrastructure-free and infrastructure-based. Infrastructure-based solutions need additional infrastructure support (e.g. ultra-wild-band (UWB), acoustic, Blue-tooth). These additional infrastructures can be used to infer range~\cite{peng2007beepbeep},pseudo-range~\cite{liu2013guoguo}, angle-of-arrival~\cite{xiong2013arraytrack}, or proximity~\cite{bahl2000radar} information to target devices. However, these approaches either require modification to end user devices, costly infrastructure or fail to achieve satisfactory positioning accuracy. In contrast, infrastructure-free solutions that use existing signal sources in indoor environments (e.g. magnetic, luminous conditions) or sensors on user devices do not require deploying additional infrastructure. Among various infrastructure-free indoor localization solutions, Pedestrian Dead Reckoning (PDR) based and Fingerprint based are the two major categories. A comprehensive survey on PDR can be found in~\cite{harle2013survey}. Most PDR-based IPS utilize the same basic modules i) step counting, ii) stride length estimation, and iii) heading estimation. In TuRF, step counting is employed to accommodate the variations in walking speed. 

In fingerprint-based IPS, although all location-dependent environmental measures can be utilized as fingerprints, magnetic field magnitudes and WiFi RSSes are most often used. Traditionally, magnetic field measurements are used for heading estimation. Magnetic field anomalies caused by building materials and magnetic interference from machinery and IT equipment make them unsuitable for heading but attractive for fingerprinting or landmark identification~\cite{li2015using,chung2011indoor}. Special devices have been developed to utilize magnetic field for small range indoor navigation~\cite{chung2011indoor}. In general, magnetic field vectors are not unique in a large area, but can be used to differentiate different locations in a small area. As a result, recent studies~\cite{li2015using,guimaraes2016motion} combine WiFi RSSes and magnetic field measurements for indoor localization. Moreover, magnetic field readings have been shown to reduce the searching space in fingerprint based indoor localization~\cite{guimaraes2016motion}.

WiFi Access Points (APs) are prevalent in indoor environments. Due to the difficulty in acquiring fine-grained synchronization and extract timing information, WiFi RSSes are more commonly used in localization. Two lines of techniques have been considered in literature: triangulation with a path loss model and fingerprinting with databases or models from site survey. Typically, WiFi RSS fingerprint-based IPS work in two stages: training and operational stages~\cite{ferris2007wifi,feng2012received,shen2013walkie,li2015using}. In the training stage, comprehensive site survey is conducted to record the fingerprints at targeted locations. In the operational stage, when a user submits a location query with her current fingerprints, a localization server computes and returns the estimated location. Existing processes for WiFi fingerprinting collection are time consuming and expensive~\cite{li2015using}. During site survey, collectors need to stand at each training position and collect WiFi scans for multiple rounds~\cite{li2015using} and possibly at different headings. In~\cite{ferris2007wifi}, the authors use the Gaussian Process Latent Variable Model (GP-LVM) to solve the WiFi SLAM problem and hence determine the latent-space locations of unlabeled signal strength data. Another relevant work is called Walkie-Markie by Shen \emph{et al.}~\cite{shen2013walkie} Walkie-Markie is an indoor pathway mapping system that can automatically reconstruct internal pathway maps of building without any a-priori knowledge about the building. Central to Walkie-Markie are the crowdsourced trajectory information (step count, step frequency, and walking direction) as well as WiFi landmarks derived from  WiFi scans. Closest to our work is the quick radio fingerprint collection (QRFC) method proposed in~\cite{liu2015quick}. In QRFC, RSS filtering and shaping are used by averaging neighboring readings along a path to compensate for signal variations along a path caused by multi-path, shadowing, and mask effects. The resulting smoothed RSSes are then stored in a database for ``nearest neighbor search" with a relational factor in the operational stage. In comparison, GP in our work performs RSS filtering and shaping automatically. Hyper-parameters in GP that control the degree of spatial smoothness and temporal variability are determined using the training data. 

Another line of work considers AP selection for localization. It was found that not all the APs contribute to indoor positioning in fingerprint-based IPS since APs have different beacon intervals and power saving mode. Previous studies show that judiciously selecting a subset of APs can improve the positioning accuracy (e.g., among those with strongest RSSes)  and 6 to 10 APs distributed around the area are often sufficient efficient~\cite{feng2012received,youssef2003wlan}. AP selection schemes are  complementary to fingerprint collection as we generally have no control over when RSSes can be collected from an AP. The former can be used during both the training and operational stages of fingerprint-based localization, and is adopted in TuRF as well. 
\section{System Overview}
\label{sect:overview}
\begin{figure}[tbp]
\centering
\includegraphics[width=0.5\textwidth]{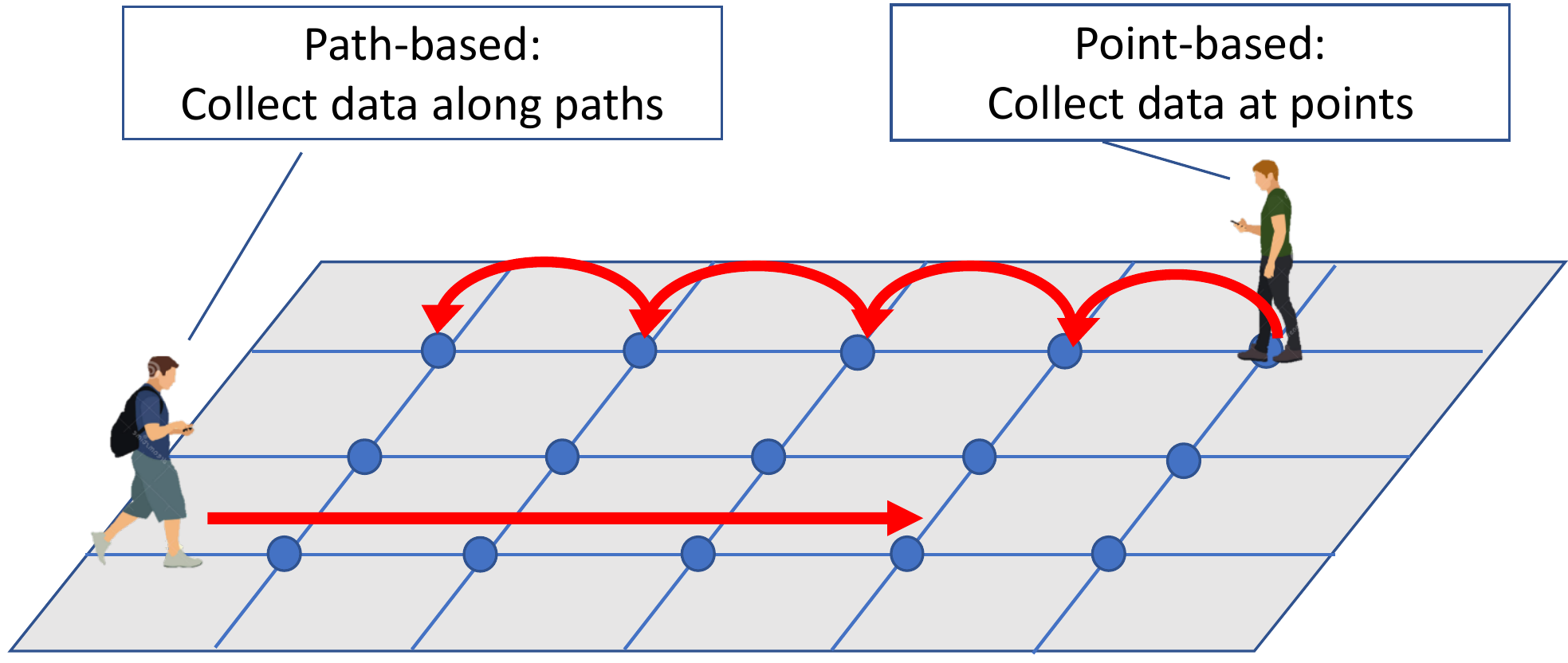}
\caption{Two data collection methods: a) collecting the fingerprints at the 15 points (indicated by  solid dots), b) collecting fingerprints along the 8 paths (indicated by thick solid lines)}
\label{fig:twomethod}
\end{figure}
The basic fingerprint-based localization problem consists of determining a device's position
${\bf v} = (x_1, x_2, ..., x_d) \in {\cal{R}}^d$, given multiple RSS
observations ${\bf{s}} = (s_1, s_2, ..., s_m) \in {\cal{R}}^m$ from $m$ APs. This is accomplished by collecting many RSSes from known locations from the target environment and train a model, also called {\it fingerprint map}, that characterizes the functional dependency between the RSS space and the coordinate space, e.g., $f_i: {\cal{R}}^d\rightarrow \cal{R}$, $i=1, ..., m$. 

Traditionally, site survey for RSS fingerprints is accomplished by first selecting a set of known locations in the target area and then collecting multiple WiFi scans while {\it standing} at each location. As an example, in Fig.~\ref{fig:twomethod}, a fingerprint map can be constructed using RSS vectors collected from the 10 intersecting grid points. TuRF instead collects both WiF RSSes and magnetic field data during walks along predefined trajectories with known starting and ending locations. In Fig.~\ref{fig:twomethod}, fingerprints are collected opportunistically along the 7 paths (indicated by thick solid lines). Several pertinent questions need to be resolved in utilizing path-based collections in training fingerprint map. 
\begin{enumerate}
\item How to infer the location tags for the unlabeled fingerprints collected during walking?
\item How fast the walks can be? 
\item How to train $f_i$'s from the fingerprints collected during walking?
\end{enumerate}

In the beginning of data collection, a set of paths are predefined. During walks, we only know exactly the locations of the starting and ending points. RSSes contained in WiFi management frames from different APs are captured opportunistically at unknown locations during the walk. In our experiments, a single WiFi scan on an Android device takes about 1s. Human step frequency is around 2Hz. Thus, multiple RSS readings can be captured during one step; and multiple steps are taken during one complete scan. 

\begin{figure}[tbp]
\centering
\includegraphics[width=0.5\textwidth]{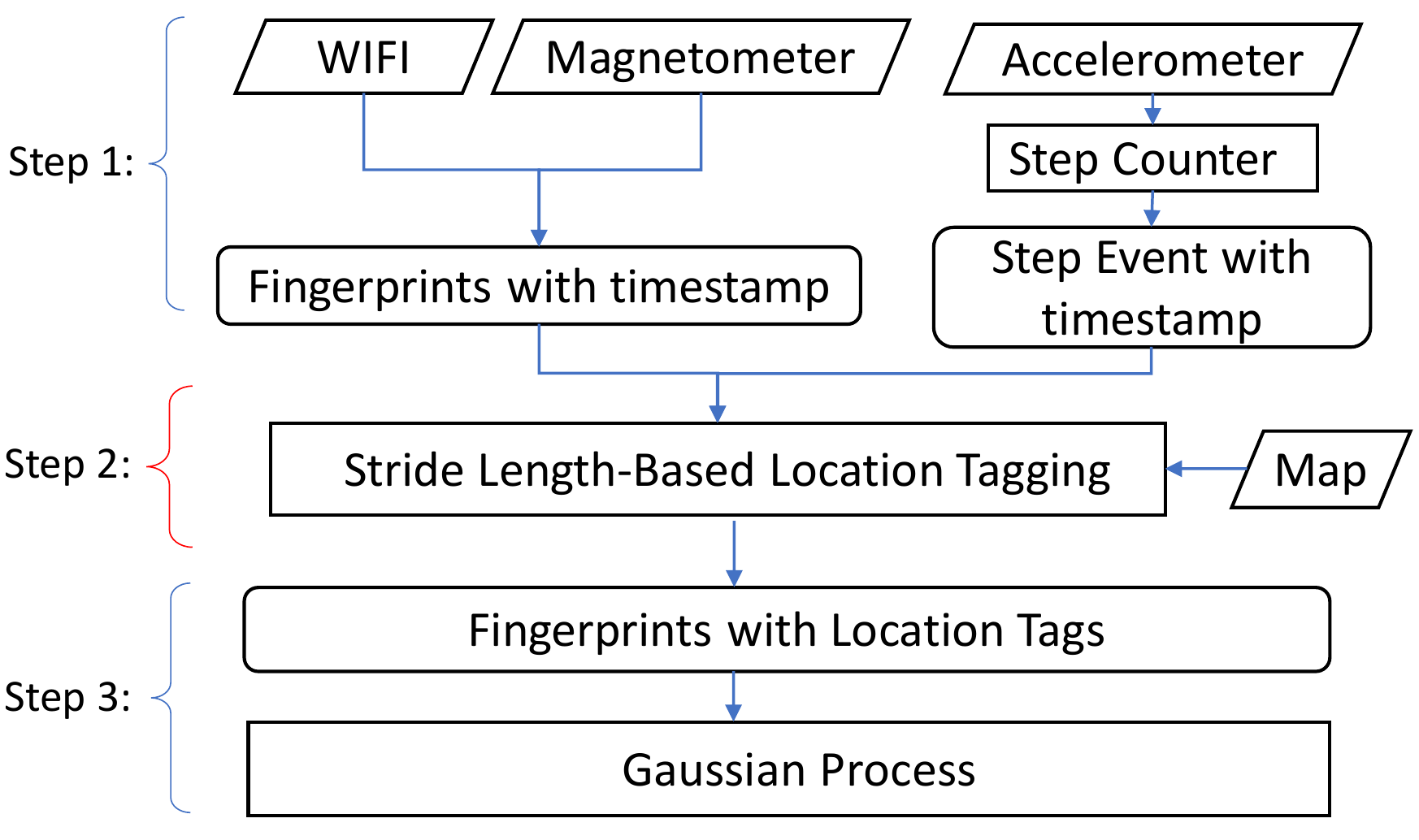}
\caption{System Architecture}
\label{fig:arch}
\end{figure}

The system architecture of TuRF is given in Fig.~\ref{fig:arch}. The proposed fast site survey process can be further divided into three steps: 1) raw data collection, 2) post processing and 3) model training. In Step 1, the WiFi interface and the magnetometer sensor on a user device are utilized to collect fingerprints with timestamps. The accelerometer sensor data are used for step detection. The timestamps of all step events are recorded. In Step 2, the step events and fingerprints are fused using their respective timestamps. A stride length based location tagging scheme is devised to assign location tags to the collected fingerprints. In Step 3, given the fingerprints with location tags, we train one Gaussian process model for each AP. 

\section{TuRF Location Tagging and Training}
\label{sect:proposed}
In this section, we present the details of TuRF location tagging and training processes. For the ease of presentation, we only consider straight-line trajectories. Paths with turns can be broken into line segments using gyroscope readings and handled similarly. 
\subsection{Location Tagging of Fingerprints}
Given a predefined path $p$ with length $L$. Its endpoints are denoted as $loc_{start}$ and $loc_{end}$. A sequence of fingerprints $\langle (fp_0, t_0), (fp_1, t_1), \cdots, (fp_n, t_n) \rangle$ are collected along this path, where $fp$ is a vector of RSS from different access points and/or magnetic filed magnitude. For any $\langle fp_i, t_i \rangle$, our goal is to infer where it was observed (i.e., its location tag). In this section, we discuss two different location inference approaches, e.g., constant speed based and constant stride length based. 

\paragraph*{Constant speed}
In order to infer the collector's location at time $t_i$ along path $p$, we can simply assume a constant walking speed. The user's location at $t_i$ can be simply calculated by
\begin{equation}
loc_i = loc_{start} + \frac{t_i - t_{start}}{t_{end} - t_{start}} \times (loc_{end} - loc_{start}), 
\label{eq:const_speed}
\end{equation}
where $t_{start}$ and $t_{end}$ indicate the moments when the user starts and stops walking on $p$ and when the operations are element-wise addition, subtraction and multiplication in the coordinate space. This approach is straightforward but is sensitive to variability in walking speed and stops during walking when encountering obstacles. In order to accommodate the speed variations during walking, we next propose a step-based algorithm.

\paragraph*{Constant stride length}
As discussed in Section~\ref{sect:overview}, we use accelerometer in the data collection process for step detection. The step events are utilized to provide better location inference. Let $L = |loc_{end} - loc_{start}|$. Given the timestamps of step events $\langle t^\prime_1, t^\prime_2, \cdots, t^\prime_K\rangle$, the constant stride length is calculated as $SL = \frac{L}{K}$. Therefore, the user's location at $t_i$ can be further inferred by
\begin{equation}
loc_i = loc_{start} + \frac{j \cdot SL + SL\frac{t_i - t^\prime_j}{t^\prime_{j+1} - t^\prime_j}}{L}\times (loc_{end} - loc_{start})
\end{equation}
where $t_j^\prime \leq t_i \leq t^\prime_{j+1}$.

\begin{figure}[tbp]
\centering
\includegraphics[width=0.5\textwidth]{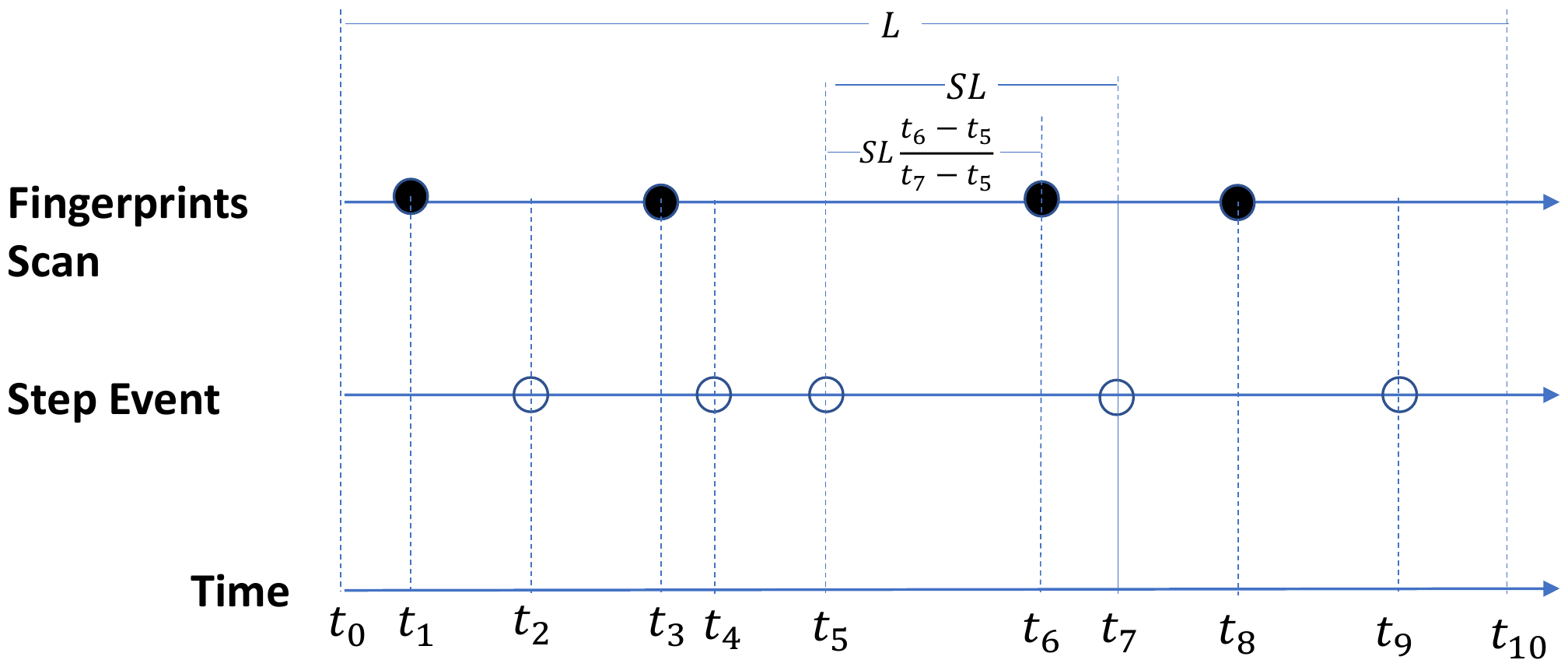}
\caption{The location tagging for unlabeled fingerprints}
\label{fig:locationtagging}
\end{figure}

Fig.~\ref{fig:locationtagging} illustrates the second approach. In the figure, there are 5 steps events and 4 fingerprints are  along the path. Therefore, the stride length is estimated $\frac{L}{5}$. Since the third fingerprint is collected during the fourth step, the location tag of the third fingerprint can be inferred as
\begin{equation}
loc_{start} + \frac{3 + \frac{t_6 - t_5}{t_7 - t_5}}{5}\times (loc_{end} - loc_{start})
\end{equation}

Obviously, a robust step counter is crucial to accurate location inference. Step counting is a core module in PDR based IPS~\cite{xuzhengubicomp15}. During one step cycle, a person's body goes through ``stance state'' with both feet on the ground, and the ``swing state'' when only one foot is on the ground. As gaits are nearly periodic, steps can be detected by identifying peak accelerations in vertical, forward or lateral directions. Let the 3-axis accelerometer readings at time $t$ be $acc_x(t), acc_y(t), acc_z(t)$. We first compute the magnitude $acc(t) = \sqrt{acc_x(t)^2+acc_y(t)^2+ acc_z(t)^2}$. Then, a low pass filter is applied to remove high-frequency components\cite{libby2012simple} from the signal. Lastly, a two-threshold based peak detection mechanism is applied to detect step event. Specially, a peak is identified as a step event if the two conditions are met, namely, i) the time difference between two adjacent steps must be greater than a chosen threshold, and ii) the magnitude difference between the adjacent peak and valley must be greater than a chosen threshold. 

\subsection{Model Training}
The outputs of site survey and location tagging are a set of fingerprints with (inferred) locations. Given a set of collected fingerprints $FP = \{(x_1, y_1), (x_2, y_2), \cdots, (x_n, y_n)\}$, where $y_i= \langle y^{bssid_0}_i, y^{bssid_1}_i, \cdots \rangle$ where $y^{bssid_j}_i$ indicates the RSS of access point $bssid_j$, observed at location $x_i$. The fingerprint-based indoor localization problem can be formally defined as: Given a set of observations $FP$, and an incoming $y^*$, how to predict $x^*$? 

\begin{table}[tbp]
\centering
\caption{The collected fingerprints by traditional point-based data collection process }
 \begin{tabular}{ | c | c | c |}
    \hline
    \textbf{Fingerprints Vector} & \textbf{Location} & \textbf{WiFi Scan ID}\\ 
    \hline
    $\langle (bssid_0^1, rss_0^1),(bssid_1^1, rss_1^1), \cdots \rangle$ & $loc_1$ & 1\\ 
    \hline
   $\langle (bssid_0^2, rss_0^2),(bssid_1^2, rss_1^2), \cdots \rangle$ & $loc_2$ & 2\\ 
   \hline
   \vdots & \vdots & \vdots\\
    \hline
    $\langle (bssid_0^N, rss_0^N),(bssid_1^N, rss_1^N), \cdots \rangle$ & $loc_N$ & N\\ 
    \hline
 \end{tabular}
 \label{tab:pointfp}
\end{table}

\begin{table}[ptb]
\centering
\caption{The collected fingerprints by path-based data collection process }
 \begin{tabular}{ | c | c | c |}
    \hline
    \textbf{Fingerprints} & \textbf{Location} & \textbf{WiFi Scan ID}\\ 
    \hline
    $(bssid_1, rss_1)$ & $loc_1$ & 1\\
    \hline
    \vdots & \vdots & \vdots\\
    \hline
    $(bssid_i, rss_i)$ & $loc_i$ & 1\\ 
   \hline
    $(bssid_{i+1}, rss_{i+1})$ & $loc_{i+1}$ & 2\\
    \hline
    \vdots & \vdots & \vdots\\
    \hline
    $(bssid_j, rss_j)$ & $loc_j$ & 2\\ 
   \hline
   \vdots & \vdots & \vdots\\
    \hline
   $(bssid_N, rss_N)$ & $loc_N$ & N\\ 
    \hline
 \end{tabular}
 \label{tab:pathfp}
\end{table}

Depending on the fingerprint collection mechanism, the size of collected fingerprints might vary. For example, the fingerprints collected by traditional point-based method is given in Table~\ref{tab:pointfp}, where a fingerprint vector consists multiple elements one from each AP observed. In contrast, the fingerprints collected by the proposed method is in fact a scalar value as illustrated in Table~\ref{tab:pathfp}. Using the data in Table~\ref{tab:pointfp} we can easily train a regression model with {\it multi-dimensional} outputs with missing values imputed~\footnote{One common way for imputing missing values in RSS vectors is to set the corresponding entry to the lowest possible readings, e.g., -93dBm}. However, this is not the case for the collected fingerprints in Table~\ref{tab:pathfp}. In this work, we assume that the RSS of different access points are independent to each other and a Gaussian process regression model will be trained separately for each access point. Formally, given $FP$ and $y^* = \langle {y^{bssid_0}}^*, {y^{bssid_1}}^*, \cdots \rangle$, $x^*$ can be estimated as

\begin{equation}
x^* = \arg\max_{x \in \mathcal{X}} \quad \prod_{i} p\left({y^{bssid_i}}^*|x\right)
\end{equation}

Now, we are in the position to discuss how to use Gaussian process to calculate the marginal likelihood $p(y|x)$. A Gaussian process can be thought of as a Gaussian distribution over functions (thinking of functions as infinitely long vectors containing the value of the function at every input). Formally, let the input space $\mathcal{X}$ and $f:\mathcal{X} \rightarrow \mathbb{R}$ a function from the input space to the reals, then we say $f$ is a Gaussian process if for any vector of inputs $\mathbf{x} = [x_1, x_2, \cdots, x_n]^T$ such that $x_i \in \mathcal{X}$ for all $i$, the vector of output $f(\mathbf{x}) = [f(x_1), f(x_2), ..., f(x_n)]^T$ is Gaussian distributed. The Gaussian process is specified by a mean function $\mu:\mathcal{X} \rightarrow \mathbb{R}$, such that $\mu(x)$ is the mean of $f(x)$ and a covariance (kernel) function $k:\mathcal{X} \times \mathcal{X} \rightarrow \mathbb{R}$ such that $k(x, x^\prime)$ is the covariance between $f(x)$ and $f(x^\prime)$. We say $f\sim GP(\mu, k)$ if for any $x_1, x_2, \cdots, x_n \in \mathcal{X}$, $[f(x_1), f(x_2),\cdots, f(x_n)]^T$ is Gaussian distributed with mean $[\mu(x_1), \mu(x_2), \cdots, \mu(x_n)]^T$ and $n\times n$ covariance/kernel matrix $K$:
\begin{equation}
K = \begin{bmatrix}
    k(x_1, x_1) & k(x_1, x_2) & \cdots & k(x_1, x_n)\\
    k(x_2, x_1) & k(x_2, x_2) & \cdots & k(x_2, x_n) \\
    \cdots & \cdots & \cdots & \cdots\\
    k(x_n, x_1) & k(x_n, x_2) & \cdots & k(x_n, x_n)
\end{bmatrix}
\end{equation}
In our model, we utilize the exponential kernel $k(x,x^\prime) = \sigma_f^2 exp(-\|x-x^\prime\|/ l)$ where $\sigma_f^2 $ is the signal variance of $f(X)$ and $l$ is the length scale of the kernel. In Gaussian process, the marginal likelihood is the integral of the likelihood times the prior
\begin{equation}
p(y|X) = \int p(y|f, X)p(f|X)df
\end{equation}
Under the zero-mean Gaussian prior assumption that $f|X \sim N(0, K)$ and the likelihood is a factorized Gaussian $y|f \sim N(f, \sigma^2_n I)$ where $I$ is an identity matrix and $\sigma_n^2$ is the noise variance, we can obtain the log marginal likelihood as
\begin{equation}
\begin{split}
log p(y|X) = & -\frac{1}{2}y^\top(K+ \sigma_n^2I)^{-1}y\\
& - \frac{1}{2}log|K+\sigma_n^2I|\\
& - \frac{n}{2}log2\pi.
\end{split}
\end{equation}
The limited-memory Broyden–Fletcher-Goldfarb-Shanno (LM-BFGS) method is employed to implement this Gaussian process regression. Please refer to~\cite{byrd1995limited} for more details. 

\subsection{WiFi RSS fingerprint selection}
Nowadays, WiFi has been deployed in almost all the indoor environments. As a result, we can always detect many access points. The effect of redundant WiFi access point is double-sided. On the one hand, ubiquitous WiFi infrastructure implies great potential for WiFi RSS based indoor localization. On the other hand, too many WiFi access points can actually compromise the localization performance, especially for those whose WiFi signals are poor. In this sense, we need a AP selection process to differentiate the ``good'' and ``bad'' access points for indoor localization. We propose a AP selection algorithm, as described in Algorithm~\ref{alg:apselect}. It works as follows: First, all the detected BSSIDs are pre-screened based on their noise variances and a noise variance threshold $\theta_{\sigma}$. Then, all the BSSIDs are further assessed based on the number of ``good'' (based on a RSS threshold $\theta_{rss}$) BSSIDs at each location and a threshold $\theta_{num}$. A BSSID might be  discarded in noise variance based pre-screening process, but it will still be retained if it is among the $\text{top-}\theta_{num}$ BSSIDs at some locations. After the BSSID selection process, a set of valid BSSIDs will be generated. In the operational stage, a user observes a set of fingerprints and submits a location query. Among the observed BSSIDs, only the valid ones will be used for localization, as elaborated in Algorithm~\ref{alg:localization}.

\begin{figure}[!t]
 \removelatexerror
    \begin{algorithm}[H]
	\caption{BSSIDSelection}
    \label{alg:apselect}
    \SetKwInOut{Input}{Input}
    \SetKwInOut{Output}{Output}
    \Input{$BSSIDs = \{ bssid_1, bssid_2, \cdots \}$, \\
    $GPRs:$ a set of GP regression models\\
    $\theta_{\sigma}: $ the threshold for noise variance,\\
    $\theta_{rss}:$ the threshold for RSS value,\\
    $\theta_{num}:$ the threshold for number of valid bssids for each position
    }
    \Output{A list of valid bssids}
    $validBSSIDs = BSSIDs$\;
    sort $BSSIDs$ based on noise variance in descending order\;
    \ForEach {$bssid_i \in BSSIDs$}
    {
    	$\sigma_n^2 = bssid_i.\sigma_n^2 $\;
        \tcc{Step 1}
    	\eIf{$\sigma_n^2 \leq \theta_{\sigma}$}{
        	isDiscarded = False\;
        }
        {
        isDiscarded = True\;
        \tcc{Step 2}
    	\ForEach{$x \in \mathcal{X}$}
        {	
        	count = 0\;
        	\ForEach{$bssid_j \in validBSSIDs$}{
            	$f = f^{bssid_j}$ \tcp*{get the GPR}
           		$rss = f(x)$\;
                
                \If{$rss \geq \theta_{rss}$}{
                	count++\;
                }
            }
            \If{$count < \theta_{num}$}{
                isDiscarded = False\;
            }
    	  }
        }
        \If{isDiscarded == True}{
        	remove $bssid_i$ from $validBSSIDs$\;
        }
    }
    	return $validBSSIDs$\;
    \end{algorithm}
    
    \begin{algorithm}[H]
	\caption{Localization}
    \label{alg:localization}
    \SetKwInOut{Input}{Input}
    \SetKwInOut{Output}{Output}
    \Input{$Observations=\{ (bssid_1, rss_1), \cdots \}$,\\
        $GPRs:$ a set of GP regression models,\\
        $validBSSIDs:$ output of {\it BSSIDSelection}
    }
    \Output{Predicted location}
		observedBSSIDs = \{bssids in $Observations$\}\;
        BSSIDs = $validBSSIDs \cap observedBSSIDs$\;
        return $ \arg\max_{x \in \mathcal{X}} \quad \prod_{bssid \in BSSIDs} p\left({y^{bssid}}|x\right)$
    \end{algorithm}
\end{figure}

\section{Performance Evaluation}
\label{sect:exp}
\begin{figure}[tbp]
\centering
\includegraphics[width=0.45\textwidth]{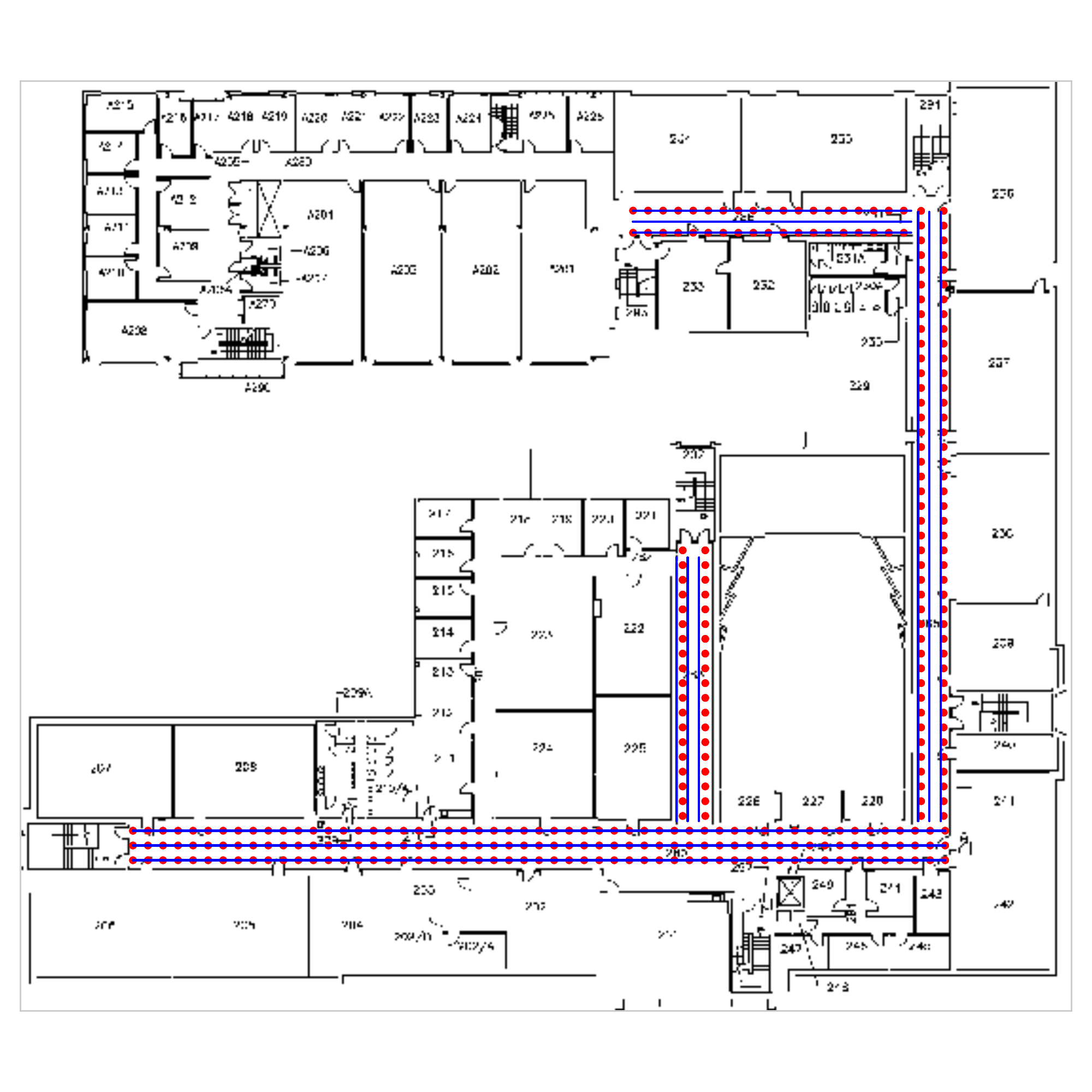}
\caption{The floor plan of the evaluation area. Red dots represent the locations selected for point-based data collection whereas the blue line segments are the predefined paths for path-based data collection}
\label{fig:itbf2}
\end{figure}

To evaluate the performance of TuRF, real world experiments are conducted on the second floor of the Information Technology Building, McMaster University. Fig.~\ref{fig:itbf2} shows the floor plan the evaluation area. The area is around $500 m^2$ with dimension 69m by 54m. Both training and testing fingerprints are collected using a Nexus 5 smart phone in the corridors. No infrastructure changes were observed during the experiments. Most of the experiments are conducted during working hours with people walking around the area. An Android App was implemented in Android for raw data collection and step detection. All data post-processing modules are implemented in Python based on open source data science libraries are used, such as Numpy, GPy, and Pandas~\cite{walt2011numpy,gpy2014,mckinney2010data}. 

In order to make WiFi scan faster, the WiFi interface on the phone is locked to 2.4 GHz using Android API. After band locking, a complete WiFi scan takes between 360ms and 420ms. In the evaluation area, there are 21 APs from our campus networks. 22 other BSSIDs are also observed during data collection in the evaluation area. After running the AP selection process described in Algorithm~\ref{alg:apselect}, a total of 33 BSSIDs are selected. In the experiments, 33 WiFi RSS features along with 2 magnetic features are utilized. As discussed in Section~\ref{sect:proposed}, a Gaussian process regression model is trained for each feature, separately.

For comparison, we collect the training data in two ways: {\it point-based data collection}, where the user stands at each of the data collection points to collect multiple scans; and the {\it path-based data collection} in TuRF, where the user walks in two directions along a set of predefined paths. The evaluation area is divided into $1.2m \times 1.2m$ grids and a total 338 points (red points in Fig.~\ref{fig:itbf2}) are selected for point-based data collection. The blue lines on Fig.~\ref{fig:itbf2} correspond to the paths for walking data collection. The distance between two parallel paths is about 1m. There are a total number of 12 predefined paths. 74 test points were selected evenly spread across the evaluation area, and testing data were collected using the point-based method. The datasets are summarized as Table~\ref{tab:dataset}.   

\begin{table}[tbp]
\centering
\caption{The description of collected fingerprints}
 \begin{tabular}{ | p{3cm} | c | c |}
    \hline
    Data Type & \# of path/point & \# of WiFi scan\\
    \hline
    Point-based Training data &338 points&10140 \\
    \hline
    Path-based Training data (Normal speed) &12 paths&2335\\
    \hline
    Test data &74 points& 4440\\
    \hline
 \end{tabular}
 \label{tab:dataset}
\end{table}

\begin{figure}[tbp]
\centering
\includegraphics[width=0.5\textwidth]{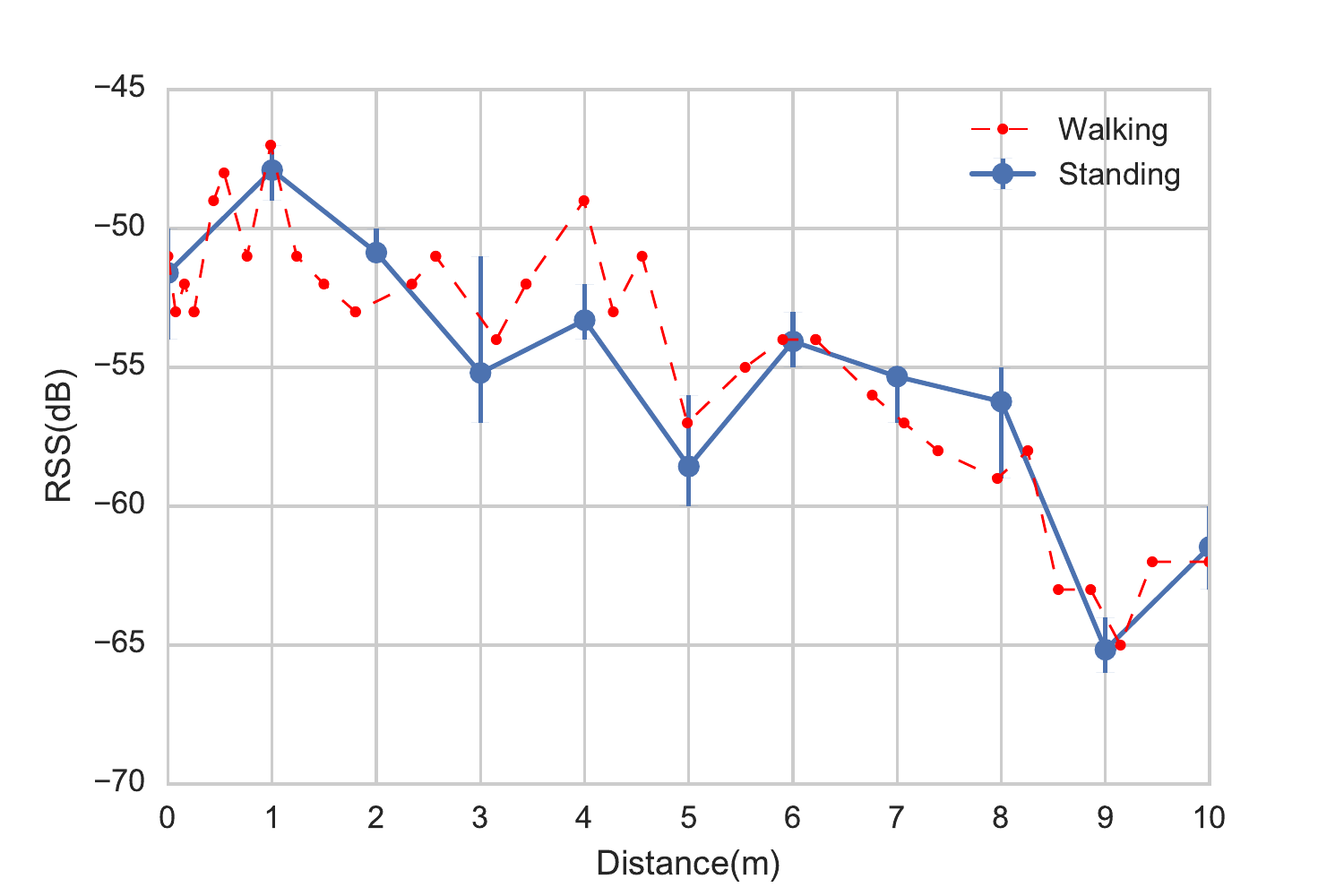}
\caption{RSS comparison between point-based and path-based data collection method}
\label{fig:walkingvsstanding}
\end{figure}

\paragraph*{RSS variations during walking}
In this experiment, we evaluate signal variations due to the shadowing effect of human body, movement and multi-path induced fading during walking. We select a 10m path and collect RSSes from the same AP using both methods. In the point-based method, RSS are collected from 11 evenly distributed locations along the path. In the path-based method, the user walks slowly along the path at roughly 0.9m/s.  Fig.~\ref{fig:walkingvsstanding} depicts the collected RSS by two different methods. The blue dots indicate the the mean RSS of 30 WiFi scans and the corresponding vertical bar represents the RSS value range (min and max) at each measuring point. The red dots represent the RSS collected during walking and the locations of these RSSes are inferred by constant stride length based location tagging discussed in Section~\ref{sect:proposed}. From Fig.~\ref{fig:walkingvsstanding}, we see that most of the RSSes observed during walking fall in the range of RSSes collected when standing. The location tags inferred from constant stride length appear to work well. Fig.~\ref{fig:walkingvsstanding} provides a clear insight as to why path-based data collection works -- it trades off scans collected per location with scans from more locations. 

\begin{figure}[tbp]
\centering
\includegraphics[width=0.40\textwidth]{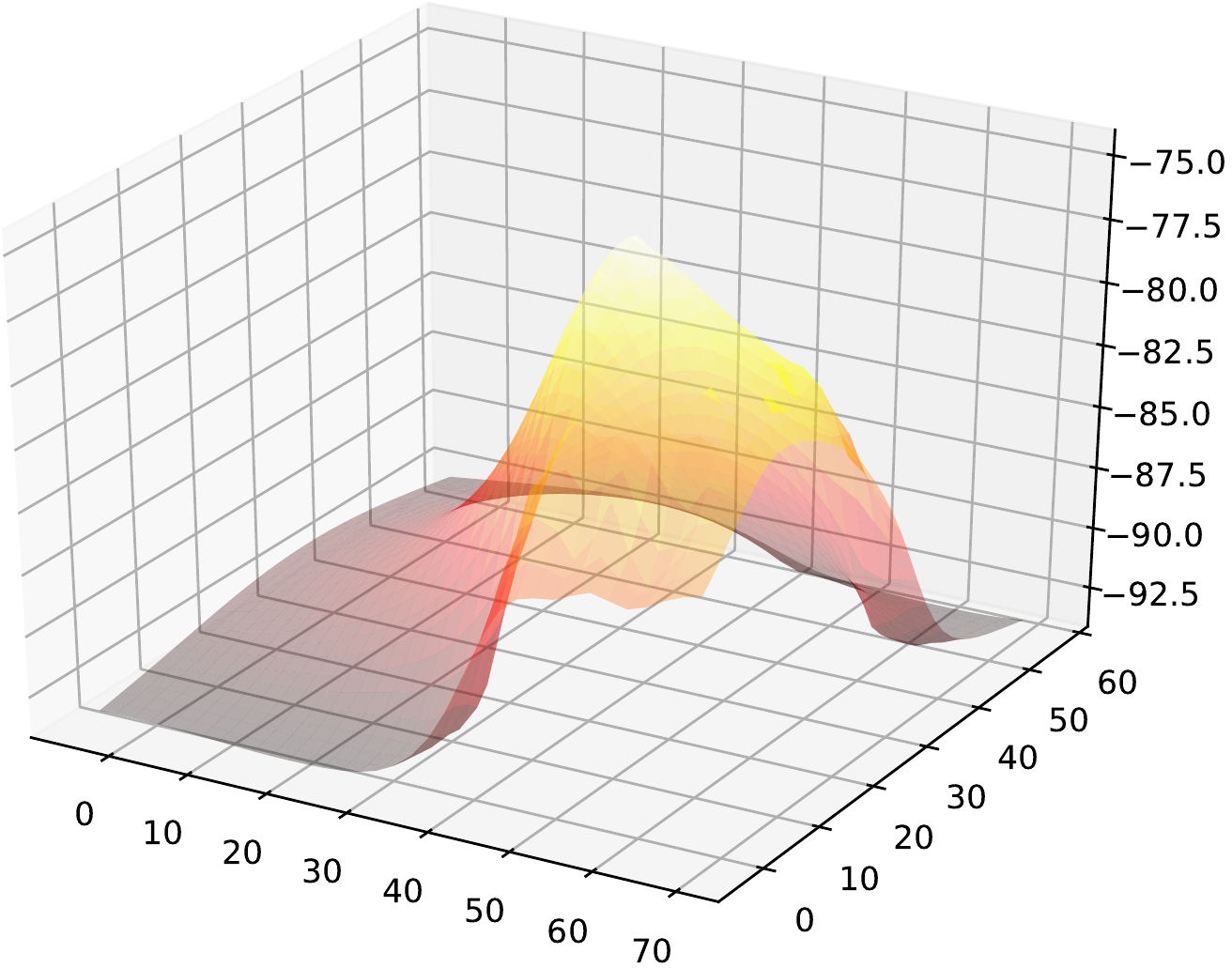}
\caption{The mean RSS value prediction using GP for one AP}
\label{fig:gp_mean}
\end{figure}

\begin{figure}[tbp]
\centering
\includegraphics[width=0.40\textwidth]{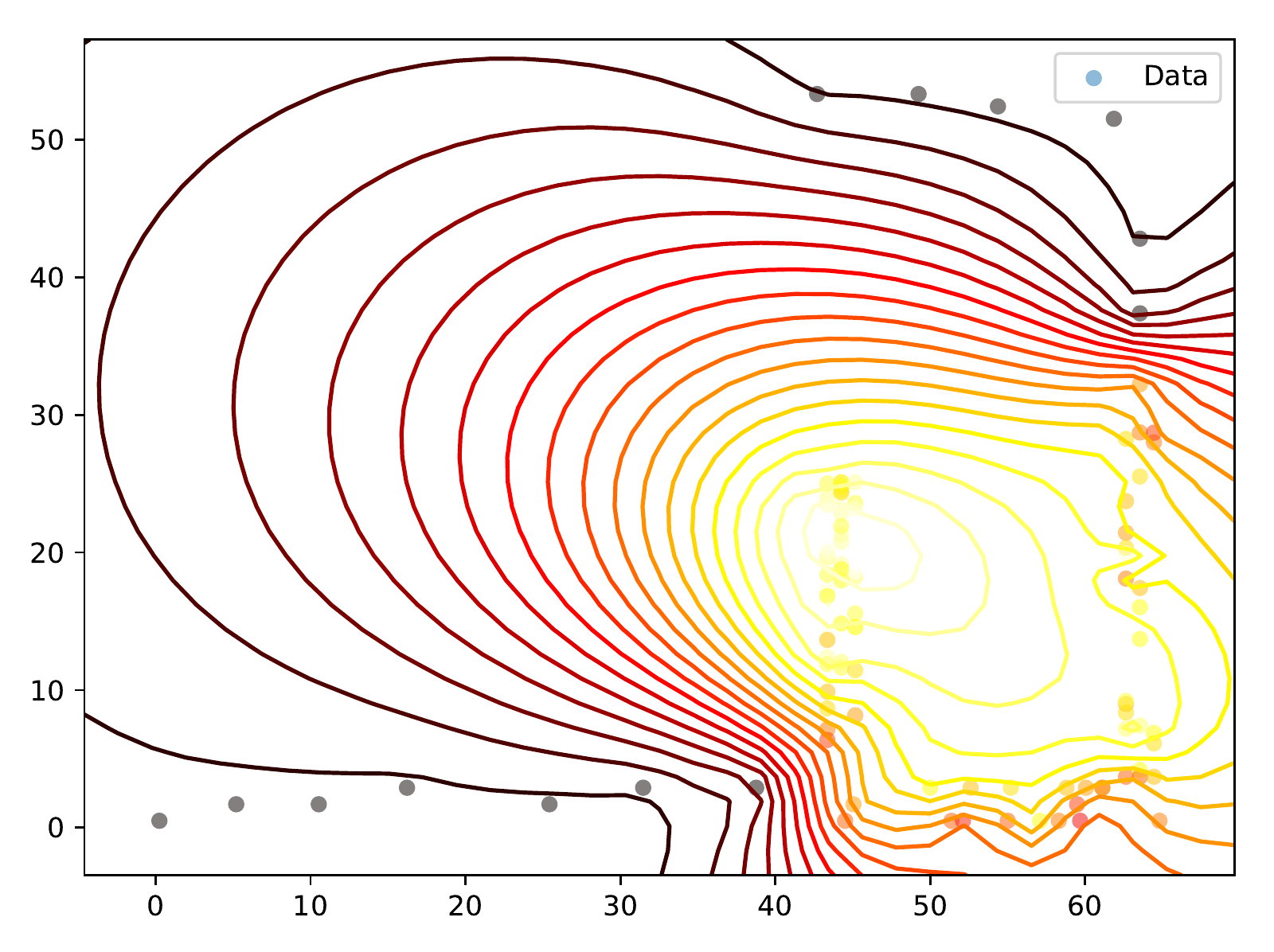}
\caption{The mean predicted RSS value and training data of one AP}
\label{fig:gp}
\end{figure}

\paragraph*{Gaussian process regression}
For each of the 35 features, one Gaussian process model was trained using the training data. The hyper parameters ($\sigma_n,\sigma_f,l$) are estimated by maximizing the marginal log-likelihood of the training data. We limit the range of noise variance $\sigma_n$ between 0.00001 and 9. The GPy framework is used in the optimization process~\cite{gpy2014}. The resulting GP generates a map with $0.1m \times 0.1m$ grids to be used for localization. The data collected from one AP during normal walking speed is shown in Fig.~\ref{fig:gp}. The colored dots indicate the data collected during walking. We impute missing data using -93 dBm if there is no RSS collected within 6 meter. Fig.~\ref{fig:gp_mean} shows the mean RSS value prediction using GP. The mean RSS value changes smoothly over the evaluation area. The predicted mean RSS value is close to the RSS observed during walking.

\paragraph*{Required time for site survey}
The total time for site survey can be broken down into i) setup time, during which markers, starting and ending points of paths are decided and measured, and ii) data collection time. Table~\ref{tab:neededtime} summarizes the setup and data collection times for different data collection methods. In TuRF, it takes us 15 minutes to measure the evaluation area and get the coordinates of the endpoints of predefined paths. For comparison, we ask the user to walk back and forth along each path at slow, normal, fast speed, respectively. It takes another 16, 27, 46 minutes to finish the data collection when the user walks at fast, normal and slow speeds, respectively. Therefore, the total times needed for data collection, when the user walks in fast, normal, and slow speed, are respectively, 31, 42, and 61 minutes. This is in contrast to data collection using the point-based method. It takes 120 minutes to measure the evaluation area and setup the markers. The total times for data collection, when 1, 10 and 30 WiFi scans are collected at each marker, are respectively, 67, 135, and 270 minutes. As shown in Table~\ref{tab:neededtime}, TuRF reduces both the setup time and the data collection time significantly. Another potential benefit of path-based data collection is that the few endpoints of paths are much easier to maintain for future fingerprint update compared to hundreds of markers needed for point-based data collection if the markers are accidentally damaged or removed. 

\begin{table}[tbp]
\centering
\caption{Time Spent on Fingerprint Collection}
 \begin{tabular}{ | p{3.8cm} | p{1.6cm} | p{2.5cm} |}
    \hline
     & Setup Time & Data Collection Time\\
    \hline
    Point-based, 1 WiFi scan at each marker&120 minutes&67minutes\\
    \hline
    Point-based, 10 WiFi scans at each marker&120 minutes&135 minutes\\
    \hline
    Point-based, 30 WiFi scans at each marker&120 minutes&270 minutes\\
    \hline
    TuRF, walk in fast speed&15 minutes & 16 minutes\\
    \hline
    TuRF, walk in normal speed&15 minutes & 27 minutes\\
    \hline
    TuRF, walk in slow speed&15 minutes & 46 minutes\\
    \hline
 \end{tabular}
 \label{tab:neededtime}
\end{table}

\begin{figure}[tbp]
\centering
\includegraphics[width=0.5\textwidth]{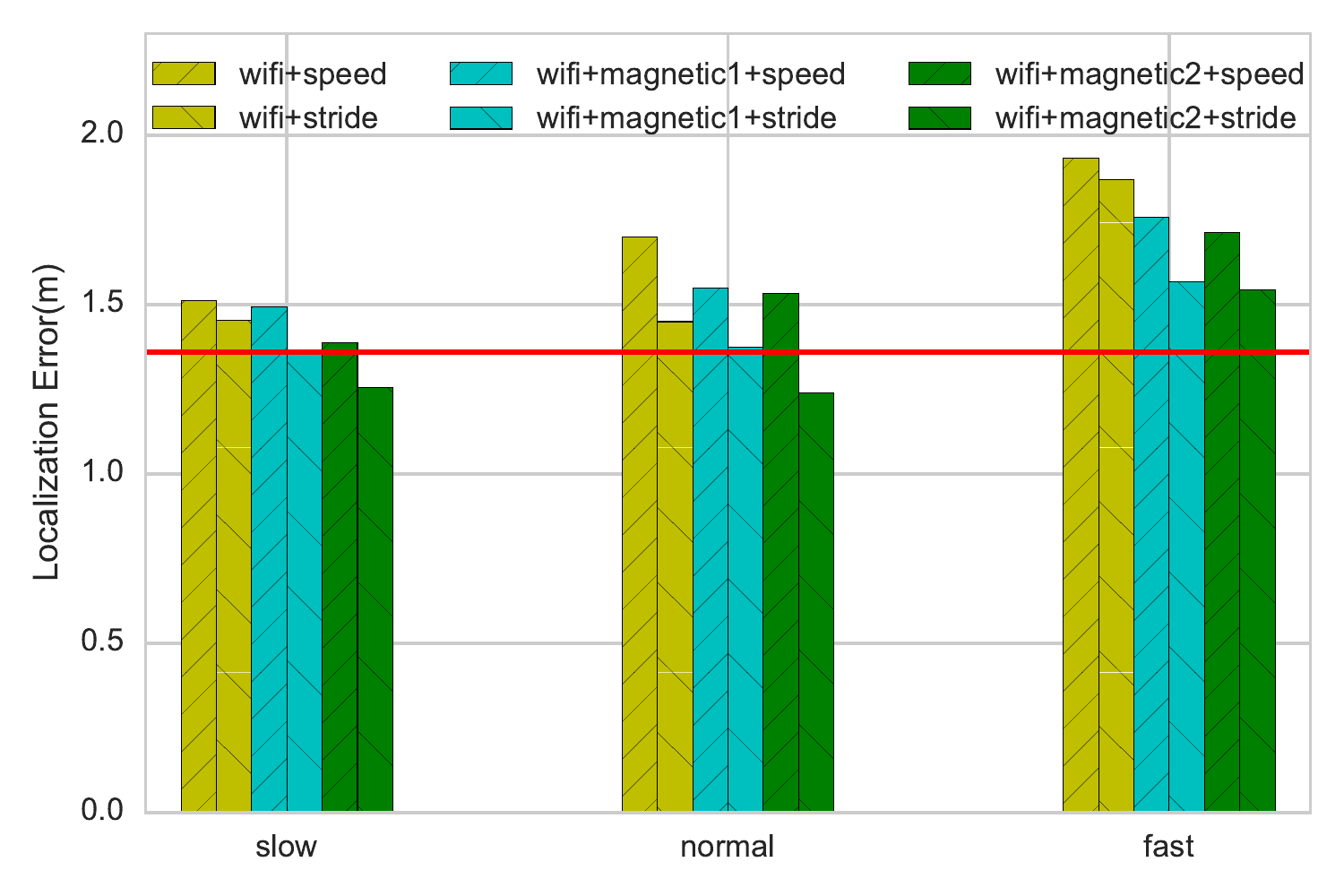}
\caption{Comparison of different strategies, walking speeds, features and location tagging approaches}
\label{fig:speedperf}
\end{figure}

\paragraph*{Localization performance}
In this experiment, we compare the localization accuracy using the training data collected by point-based data collection and TuRF. We also evaluate the impacts of location tag inference methods, walking speed and magnetic features. Six variants of the combined techniques are evaluated.
\begin{itemize}
\item {\it WiFi+speed:} Only WiFi RSS fingerprints are used. The location tagging is based on constant speed.

\item {\it WiFi+stride:} Only WiFi RSS fingerprints are used. The location tagging is based on constant stride length.

\item {\it WiFi+magnetic1+speed:} Both WiFi RSS and magnetic field magnitude are used as fingerprints. The location tagging is based on constant speed.

\item {\it WiFi+magnetic1+stride:} Both WiFi RSS and magnetic field magnitude are used as fingerprints. The location tagging is based on constant stride length.

\item {\it WiFi+magnetic2+speed:} Both WiFi RSS, magnetic field magnitude and magnetic field magnitude on Z-axis are used as fingerprints. The location tagging is based on constant speed.

\item {\it WiFi+magnetic2+stride:} Both WiFi RSS, magnetic field magnitude and magnetic field magnitude on Z-axis are used as fingerprints. The location tagging is based on constant stride length.
\end{itemize}
The results are depicted as Fig.~\ref{fig:speedperf}. The red horizontal line indicates the performance of  point-based data collection algorithm in which both WiFi and magnetic fingerprints are utilized. The vertical bars represent the localization performances of path-based data collection methods. A few observations are in order from Fig.~\ref{fig:speedperf}. First, localization results from the constant stride length based location tagging algorithm are constantly better than those from the constant speed based algorithm in all the cases. This indirectly confirms that location tags using constant stride length are likely to be more accurate. Second, incorporation of magnetic fingerprints can indeed improve the localization performance. Using both magnetic features are beneficial. Third, when constant stride length location tagging is used, TuRF outperforms point-based data collection when the walking speed is slow and normal. Recall from Table~\ref{tab:dataset}, the total numbers of WiFi scans in the point-based method and TuRF at the normal walking speed are 10140 and 2335, respectively. With fewer scans, the superior performance of TuRF can be attributed to denser spatial sampling. Fourth, as expected, slower walking speeds allow collection of fingerprints and thus lead to better localization accuracy. 

\paragraph*{Walking speed recommendation}
As shown in Fig.~\ref{fig:speedperf}, we observe that lower walking speed leads to better localization performance in TuRF. However, there exists a trade-off between the time spent on site survey and localization accuracy. Through extensive experiments, we find that a reasonable localization performance can be attained if there are one or more WiFi scans that can be done within one step. Based on this observation, the recommended walking speed is given by, 
\begin{equation}
recommended\ walking\ speed \leq \frac{1}{t_{WiFi Scan}}
\end{equation}

\begin{table}[tbp]
\centering
\caption{Walking speed recommendation for different devices}
 \begin{tabular}{ | c | p{2cm} | p{2cm} |}
    \hline
    Device & Required time for one WiFi Scan & Recommended walking speed\\
    \hline
    Nexus 5 & 380 ms & 2.63 step/s\\
    \hline
    Samsung Galaxy Note 3 & 1420 ms& 0.70 step/s\\
    \hline
    Samsung Galaxy S4 & 780 ms& 1.28 step/s\\
    \hline
    Samsung Galaxy Mini III & 1150 ms & 0.87 step/s\\
    \hline
 \end{tabular}
 \label{tab:recommendation}
\end{table}
The amount of time for one WiFi scan is both device and configuration dependent. Rule-of-thumb walking speed recommendations for different Android devices are given in Table~\ref{tab:recommendation}.

\section{Conclusion}
\label{sect:conclude}
In this paper, we presented TuRF, a fast path-based data collection method for fingerprint collection. The Gaussian process regression model was utilized to provide a flexible model training process for localization. Experimental results demonstrated the efficiency and effectiveness of the proposed method. We found that TuRF can indeed reduce the required time for site survey without sacrificing localization performance. As future work, we are interested in incremental fingerprints updating strategies and investigate the adaptive data collection approaches where users are promoted if walk speeds shall be adjusted and revisits to some areas are required.

\balance
\bibliographystyle{IEEEtran}
\bibliography{ref}
\end{document}